\documentclass[
  ,draft            
  ]
  {aipproc}

\usepackage{amsfonts}
\usepackage{amssymb}
\usepackage{amsmath}
\usepackage{amsbsy}
\usepackage{graphicx}
\usepackage{natbib}
\usepackage[english]{babel}

\newcommand{\sech}{{\rm sech\,}}

\newcommand{\lunit}{\ensuremath{\,\mbox{km}}}
\newcommand{\vunit}{\ensuremath{\,\mbox{km} \, \mbox{s}^{-1}}}

\newcommand{\nunit}{\ensuremath{\,\mbox{cm}^{-3}}}
\newcommand{\tunit}{\ensuremath{\,\mbox{s}}}

\newcommand{\ba}{\begin{eqnarray}}
\newcommand{\ea}{\end{eqnarray}}

\newcommand{\bas}{\begin{eqnarray*}}
\newcommand{\eas}{\end{eqnarray*}}

\newcommand{\be}{\begin{equation}}
\newcommand{\ee}{\end{equation}}

\newcommand{\bes}{\begin{equation*}}
\newcommand{\ees}{\end{equation*}}

\layoutstyle{8x11double}

\begin{document}

\title[Energy release into the solar atmosphere from magnetic reconnection]{Analysis of the energy release for different magnetic reconnection regimes within the solar environment}


\author{Lapo Bettarini}{
  address={Centrum voor Plasma-Astrofysica, Departement Wiskunde, Katholieke Universiteit Leuven, Celestijnenlaan 200B, B-3001 Leuven, Belgium}
}

\author{Giovanni lapenta}{
  address={Centrum voor Plasma-Astrofysica, Departement Wiskunde, Katholieke Universiteit Leuven, Celestijnenlaan 200B, B-3001 Leuven, Belgium}
}

\begin{abstract}
A 2.5-dimensional magnetohydrodynamics simulation analysis of the energy release for three different reconnection regimes is presented. The system under investigation consists in a current-sheet located in a medium with a strong density variation along the current layer: such system is modeled as it were located in the high chromosphere/low solar corona as in the case of pre- flare and coronal mass ejection (CME) configurations or in the aftermath of such explosive phenomena. By triggering different magnetic-reconnection dynamics, that is from a laminar slow evolution to a spontaneous non-steady turbulent reconnection~\citep{lapenta08,bettarini09,skender09}, we observe a rather different efficiency and temporal behavior with regard to the energy fluxes associated with each of these reconnection-driven evolutions. These discrepancies are fundamental key-properties to create realistic models of the triggering mechanisms and initial evolution of all those phenomena requiring fast (and high power) magnetic reconnection events within the solar environment.
\end{abstract}

\maketitle


\section{Introduction}
Magnetic reconnection is essential to model partially or completely most of the phenomena occurring within the solar atmosphere, being the sun's magnetic field the huge reservoir of energy driving the dynamical evolution of such structures. A detailed and realistic model of magnetohydrodynamics (MHD) reconnection is an unescapable step in order to achieve a deep understanding of the highly dynamic solar atmosphere as well as of the interaction of solar plasma structures with the interplanetary medium. In fact, the MHD approach is adequate to describe the initial evolution and the following nonlinear stages of such phenomena, even though the detailed physics of their initiation depends primarily on kinetic effects. 

In general, the starting point is either the well-known Sweet-Parker's model of a slow and laminar reconnection~\citep{sweet58,parker57} in a harris or force-free current-sheet, or the fast process described by Petschek's model~\citep{petschek64} and usually obtained by the introduction in the MHD system of a localized enhancement of the plasma resistivity~\citep{biskamp00,yokoyama01}. Yet, a new high-power and non-steady turbulent reconnection process is possible in MHD without invoking the presence of any \textit{ad-hoc} term resembling kinetic effects in MHD equations. This has been recently shown by~\citet{lapenta08} and applied to the solar environment by~\citet{bettarini09}. Such mechanism includes a spontaneous transition from a slow to a fast reconnection with time-scales of the order of Alfv\'en times and no factitious anomalous effects are considered. During its evolution, current-sheets develop a chaotic structure of multiple small-scale interacting reconnection sites determining a macroscopic turbulent state.

As in~\citet{bettarini09}, the system under investigation in the present work is a current-sheet embedded in the high chromosphere/low solar corona region. It consists in a current-sheet determined by a force-free magnetic field in the presence of a strong density variation along the current layer provided by a proper step function. We vary the global resistivity of our numerical domain (by means of the Lundquist number) such that different dynamical regimes of the current-sheet are set, from a slow and diffusive reconnection to a fast and turbulent process and an in-between evolution. Here, we present an energy-release analysis of such three different processes in order to (a) provide useful hints of the best-promising mechanism to be considered for more and more realistic models of solar (explosive) phenomena, and (b) to determine the key-features of events that could be observed by current and near-future solar and space missions.

\section{Experiment Settings}

We solve numerically the viscous-resistive compressible $2.5$D one-fluid MHD equations parametrized by the global Lundquist and kinetic Reynolds number, {\bf S} and {\bf R}$_\nu$ respectively. We define a vertical or stream-wise direction ($z$) oriented away from the sun, wherein open boundary conditions are set at $z = 0$ and $L_z$, and a horizontal or cross-stream direction ($x$) where we set reflecting boundary conditions. We consider an ideal equation of state with polytropic index $\gamma$ such that $p = \rho \, (\gamma-1) \, I$ where $p$, $\rho$ and $I$ are respectively the kinetic pressure, the density and the enthalpy of the plasma. We use the Lagrangian code, FLIP3D-MHD~\citep{brackbill91}, and we consider $600$ (in $x$) $\times$ $960$ (in $z$) Lagrangian markers arrayed initially in a $3 \times 3$ uniform formation in each of the $200 \times 320$ cells of our numerical grid.
\begin{table}
\begin{tabular}{lccc}
\hline
    & \tablehead{1}{c}{b}{Run A} & \tablehead{1}{c}{b}{Run B} & \tablehead{1}{c}{b}{Run C} \\
\hline
S & $10^2$ & $10^3$ & $10^4$ \\
\hline
\multicolumn{4}{c}{$\beta = 0.2$, R$_\nu = 10^4$, $\gamma = 5/3$} \\
\multicolumn{4}{c}{$x \in [-10.,10.]$, $z \in [0.,80.]$} \\
\multicolumn{4}{c}{$dx = 0.1$, $dz = 0.25$} \\
\hline
\end{tabular}
\caption{Summary of the simulations performed for the present work and the related parameters.}
\label{table1}
\end{table}
As already pointed out in the introduction, our system consists in a current-sheet configuration determined by a force-free magnetic field with $B_{z}(x) = \tanh x$, $B_{x}(x) = 0$ and the out-of-plane component of the magnetic field $B_{y}(x) = -\sech x$ as guide field. With this configuration, we obtain high-resolution simulations where the current-sheet width is resolved by $30$ Lagrangian markers. The density is modeled as a step function where the ratio between the density in the region $0 < z < L_z/8$ (modeling the high chromosphere) and that one in $L_z/8 < 0 < L_z$ (the low corona) is about $10^5$ as considered in~\citet{yokoyama01, bettarini09}. We consider an initial GEM-like perturbation~\citep{birn01} but exponentially localized at the center of the numerical box (i.e. within the solar corona) with a perturbation amplitude of $\epsilon = 0.5$. 

\subsection{Physical Parameters} 

The MHD equations are normalized according to reasonable values for the high chromosphere/low solar corona: the number density $\rho/m \approx 10^9 \nunit$; the length scale $L \approx 1000-2000 \lunit$, where $L$ is the current-sheet width and $L = L_x/20 = L_z/80$; the Alfv\'en velocity $v_A \approx 400-600 \vunit$ (approximately three times the sound speed). So, we have a time scale of $t_A = L/v_A \approx 1.5-5 \tunit$. See Tab.~\ref{table1} for more details on simulations' setting and parameters.

\section{Results} 
\begin{figure}[b!]
\includegraphics[height=.25\textheight]{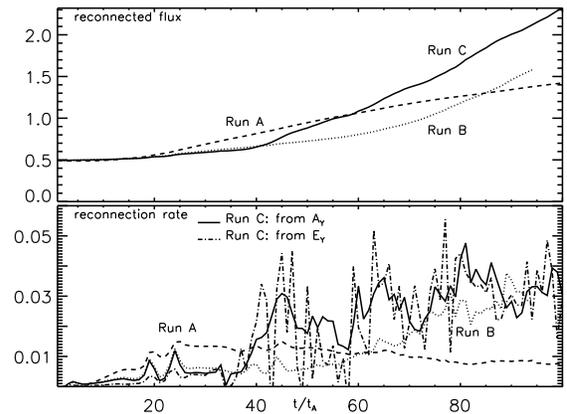}
\caption{Top panel: Measured reconnected flux as a function of time where the dashed line corresponds to Run A, dotted line to Run B and solid and dotted-dashed lines correspond both to Run C (see Tab.~\ref{table1}). Bottom panel: Reconnection rate as a function of time for the three simulations.}
\label{figure1}
\end{figure}
In the top panel of Fig.~\ref{figure1}, the reconnected magnetic flux (in dimensionless units) as a function of time for three different values of the Lundquist number (so, of the global resistivity) is shown. This quantity is measured by computing the maximum and minimum values of the out-of-plane magnetic potential, $A_y$ at the mid axis ($x = 0$): their difference provides the amount of flux in the closed field lines, caused by reconnection as described in~\citet{biskamp00}. A low value of {\bf S}, {\bf S}$ = 10^2$ results in a laminar and slow dynamics of the current-sheet. After an initial transient time, a nearly-constant reconnected flux is observed (Run A, dashed line). By setting {\bf S} to $10^3$, the system presents again an initial nearly-constant increase in the reconnected flux resembling a slow reconnecting evolution, although it is  more and more evident a smooth transition to a faster evolution as the time goes by (Run B, dotted line). By increasing {\bf S} further, {\bf S} $=10^4$, smaller scales can be reached and the high resolution allows us to observe a completely different regime~\citep{lapenta08,bettarini09}. Because of a tearing instability of the initial Sweet-Parker configuration~\citep{furth63,skender09}, the system experiences an abrupt switch from an initial slow and laminar reconnection regime to a fast reconnection process leading to a macroscopic turbulent configuration of the initial system. 
\begin{figure*}
\includegraphics[height=.3\textheight]{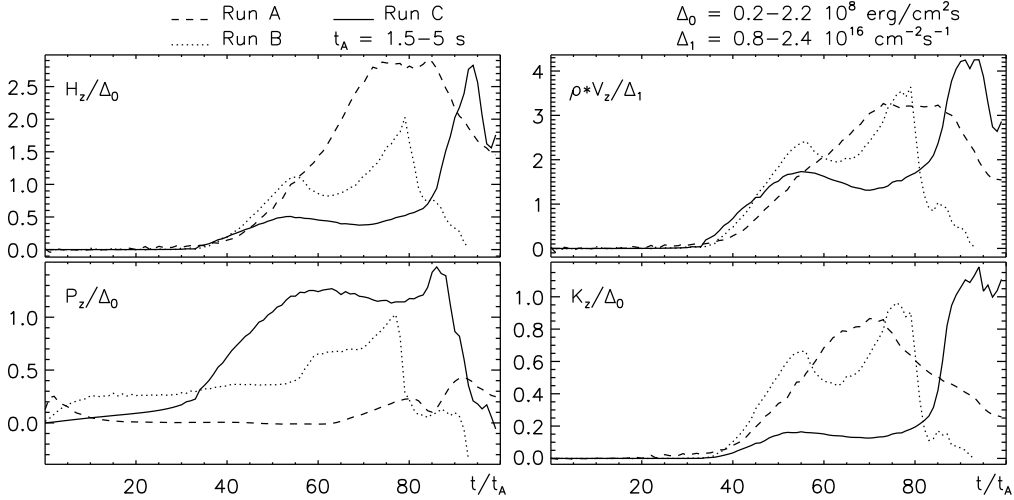}
\caption{Fluxes through the upper boundary of the numerical box: enthalpy (top-left panel), mass (top-right panel), Poynting vector (bottom-left panel), and kinetic energy (bottom-right panel).}
\label{figure2}
\end{figure*}
The different behavior of the reconnection process in the three simulations is pointed out also in the evolution of the reconnection rate as a function of time. This is shown in the bottom panel of Fig.~\ref{figure1}. Different rates but the same general behavior are shown during the first few decades of time steps where the system is driven by a slow reconnection. At about $t = 40$ (measured in terms of alfv\'en times), it is evident the rapid increase of the reconnection rate in Run C. With regard to this case, the rate is measured both by computing the slope of the corresponding reconnected flux (solid line in~Fig.~\ref{figure1}) as for the other runs and also by means of the out-of-plane electric field variation through the reconnection diffusion region (dotted-dashed line). With the latter method, the curve of the rate is more noisy, yet the two methods give two manifestly consistent results. Note that also the rate of Run B (dotted line) is becoming more and more consistent with a fast evolution as the time goes by, whereas the Run A's rate is slowly decreasing. 

As described in \citet{bettarini09}, during such three different regimes, the slow reconnection evolution, the spontaneous and abrupt variation from a slow to a fast regime and the intermediate dynamics, the current-layer develops asymmetrically because of the strong density variation resembling the transition from the high chromosphere and the low solar corona. Moreover, characteristic features in magnetic field topology of the aftermath of solar explosive phenomena are observed. In order to understand the importance of such reconnection mechanisms within the solar atmosphere, a fundamental step is to analyze such processes from the point of view of their efficiency in releasing energy and particles in the upper atmosphere. In Fig.~\ref{figure2}, the flux of the Poynting vector, the enthalpy flux, the kinetic flux, and the mass flux through the upper boundary of the numerical box ($z = L_z$) are shown as functions of time. In the figure also the normalization values for all the shown quantities are reported. As the perturbation is approximately in the middle of our numerical box ($z \sim L_z/2 = 40$, that is in the solar corona) and there the alfv\'en velocity is $1$, we start to measure the effects of reconnection at the upper boundary from $t \sim 40$. Of course, we neglect the initial transient time caused by a perturbation that is not exactly an eigenstate of the system. 

In the Run A, the current-sheet evolves accordingly a diffusive reconnection process that determines a pile-up of the enthalpy flux of the system (top-left panel of Fig.~\ref{figure2}) as well as an increase of the mass flux (top-right panel) to the upper layers of the corona because of an acceleration process as shown by the kinetic flux's profile (bottom-right panel of Fig.~\ref{figure2}). The flux through the upper boundary of the stream-wise component of the Poynting vector is almost null throughout the simulation. As previously pointed-out, Run A's reconnection rate sensitively decreases in the last phase and after a plateau of a few time steps, we observe a corresponding decrease of all fluxes. 

The slow-to-fast reconnection evolution of the current-sheet observed in Run C passes through a chaotic state during which the formation of several smaller-scale current-sheets and reconnection sites interacting each other is observed~\citep{bettarini09}. So, it is the interaction of such structures continuously moving backward and forward that prevents the system to release a large amount of energy and particles to the upper atmosphere. Yet, as the process goes on and the system reaches a macroscopic turbulent configuration, an impulsive event is observed and a release of mass as well as an increase of the temperature is rapidly determined in the upper corona. A different consideration has to be done with regard to Poynting vector's flux. In fact, it increases constantly and sensitively during all the fast regime, but it does not characterize the impulsive phase of the system.

As already pointed out, an intermediate evolution is observed during the Run B. The system switches to a fast reconnection regime according to a slow rather than an abrupt process. From the energy- and mass-flux point of view, the profiles shown in~\ref{figure2} stay in-between the behavior described for the previous cases. An initial slow but constant increase of the fluxes is followed by a plateau of few time-steps that does not lead to a final decrease of the fluxes as in the Run A but to a impulsive phase as in the Run C, though being the peak values sensitively lower. 	

\section{Conclusions}

Here, we observe the energetic and mass release of a force-free current-sheet evolving in a small region of the high/chromosphere/low solar corona. According to the new mechanism proposed by~\citet{lapenta08} and applied by~\citet{bettarini09} within the solar environment, it is possible that this system follows a magnetic field-line reconnection dynamics and that spontaneously evolves from a low- and constant-rate phase to a fast and high power regime within the pure-resistive MHD-theory framework (and so without invoking the presence of any anomalous resistivity). Hence, it becomes fundamental the understanding whether this process can provide us with evolution-related observables that can improve our models of solar (explosive) phenomena and of MHD structures in general.

By performing 2.5D MHD high-resolution simulations and by varying the characteristic global Lundquist number, we switch three different magnetic reconnection regimes: a Sweet-Parker evolution, the new spontaneously slow-to-fast reconnection and an intermediate dynamics. By means of the Sweet-Parker regime, we can obtain an quantitatively efficient but slow conversion of the initial magnetic energy available in the basic configuration, but it does not provide some important features observed for example in solar phenomena like flares or CMEs, that is a spontaneous impulsive event that evolves on time-scales of tens/hundreds of seconds (see again Fig.~\ref{figure2} and the normalization values there reported) and releases large quantities of mass, enthalpy and high-velocity structures that eventually can power and trigger secondary events in the upper solar atmosphere. Even an intermediate evolution combining an initial laminar reconnection dynamics and evolving to a faster and faster regime is not able to recover such dynamical features, even if a less rapid and power impulsive process is observed. Here, we show that the spontaneous non-steady reconnection process can provide time-scales, reconnection rates and energy- and mass-flux values that, on the one hand, are consistent with the theoretical framework of pure resistive-MHD and, on the other hand, they allow to use such mechanism as starting model of the complex triggering event and initial evolution of solar phenomena. 

Future studies will include a detailed analysis both from the dynamical and energetic point of view of the smaller-scale structures forming during the chaotic evolution of the current-sheet by means of adaptive-mesh refinement numerical tools. Furthermore, at this stage the introduction of a more realistic model of the high chromosphere/low solar corona in a three-dimensional configuration is mandatory.


\begin{theacknowledgments}
The research leading to these results has received funding from the European Commission's Seventh Framework Programme (FP7/2007-2013) under the grant agreement n$^\circ$ 218816 (SOTERIA project,  \url{www.soteria-space.eu}. The simulations shown were conducted using processors on the VIC cluster of the Vlaams Supercomputer Centrum at K.U. Leuven (Belgium). 
\end{theacknowledgments}


\begin{thebibliography}{11}
\expandafter\ifx\csname natexlab\endcsname\relax\def\natexlab#1{#1}\fi
\providecommand{\enquote}[1]{``#1''}
\expandafter\ifx\csname url\endcsname\relax
  \def\url#1{\texttt{#1}}\fi
\expandafter\ifx\csname urlprefix\endcsname\relax\def\urlprefix{URL }\fi
\providecommand{\eprint}[2][]{\url{#2}}

\bibitem[Lapenta(2008)]{lapenta08}
G.~Lapenta, \emph{Phys. Rev. Lett.} \textbf{100}, 235001 (2008).

\bibitem[Bettarini and Lapenta(2009)]{bettarini09}
L.~Bettarini, and G.~Lapenta, \emph{ApJ} \textbf{Submitted} (2009).

\bibitem[Skender and Lapenta(2009)]{skender09}
M.~Skender, and G.~Lapenta, \emph{Phys. Plasmas} \textbf{submitted} (2009).

\bibitem[Sweet(1958)]{sweet58}
P.~A. Sweet, \enquote{The Neutral Point Theory of Solar Flares,} in
  \emph{Electromagnetic Phenomena in Cosmical Physics}, edited by B.~Lehnert,
  Cambridge University Press, Cambridge, England, 1958, vol.~6 of \emph{IAU
  Symposium}, pp. 123--+.

\bibitem[Parker(1957)]{parker57}
E.~N. Parker, \emph{J. Geophys. Res.} \textbf{62}, 509--520 (1957).

\bibitem[Petschek(1964)]{petschek64}
H.~E. Petschek, \enquote{Magnetic field annihilation,} in \emph{The Physics of
  Solar Flares}, edited by W.~N. Hess, 1964, pp. 425--+.

\bibitem[Biskamp(2000)]{biskamp00}
D.~Biskamp, \emph{Magnetic reconnection in plasmas}, Cambridge Monograph on
  Plasma Physics, Cambridge University Press, Cambridge, England, 2000.

\bibitem[Yokoyama and Shibata(2001)]{yokoyama01}
T.~Yokoyama, and K.~Shibata, \emph{ApJ} \textbf{549}, 1160--1174 (2001).

\bibitem[Brackbill(1991)]{brackbill91}
J.~U. Brackbill, \emph{J. Comp. Phys.} \textbf{96}, 163--192 (1991),
  \urlprefix\url{http://adsabs.harvard.edu/abs/1991JCoPh..96..163B}.

\bibitem[Birn et~al.(2001)]{birn01}
J.~Birn, J.~F. Drake, M.~A. Shay, B.~N. Rogers, R.~E. Denton, M.~Hesse,
  M.~Kuznetsova, Z.~W. Ma, A.~Bhattacharjee, A.~Otto, and P.~L. Pritchett,
  \emph{J. Geophys. Res.} \textbf{106}, 3715--3720 (2001).

\bibitem[{Furth} et~al.(1963)]{furth63}
H.~P. {Furth}, J.~{Killeen}, and M.~N. {Rosenbluth}, \emph{Phys. Fluids}
  \textbf{6}, 459--484 (1963).

\end{thebibliography}
\end{document}